\documentclass[12pt]{article}
\usepackage{epsfig}

\def\beq{\begin{equation}}
\def\eeq#1{\label{#1}\end{equation}}
\def\eeqn{\end{equation}}

\def\beqa{\begin{eqnarray}}
\def\eeqa#1{\label{#1}\end{eqnarray}}
\def\eeqan{\end{eqnarray}}

\let\bar=\overbar

\def\Dslash{\not{\hbox{\kern-4pt $D$}}}
\def\dslash{\not{\hbox{\kern-2pt $\del$}}}

\def\msb{{\bar{\ssstyle M \kern -1pt S}}}

\newcommand{\dzero}  {D0}
\newcommand{\ttbar}  {\mbox{$t\bar{t}$}}
\newcommand{\pt}     {\ensuremath{p_{\rm T}}}
\newcommand{\met}    {\mbox{$\not\!\!E_T$}}
\newcommand{\vtb}    {\ensuremath{V_{\rm tb}}}
\newcommand{\pb}     {\ensuremath{{\rm pb}^{-1}}}
\newcommand{\fb}     {\ensuremath{{\rm fb}^{-1}}}

\def\Title#1{\begin{center} {\Large {\bf #1} } \end{center}}

\begin{document}

\Title{Top Quark Cross Sections}

\bigskip\bigskip


\begin{raggedright}  

{\it Fr\'ed\'eric D\'eliot\index{Deliot, F.}\\
CEA-Saclay, IRFU/SPP, b\^at. 141, 91191 Gif sur Yvette Cedex, France\\
Frederic.Deliot@cea.fr\\
On Behalf of the CDF and DO Collaborations.}
\bigskip\bigskip
\end{raggedright}

\section{Introduction}

The top quark is the latest quark of the Standard Model (SM) that was
discovered, discovery that was made in 1995 by the CDF and \dzero\ 
collaborations~\cite{topdiscovery}. It is the ``youngest'' quark of the SM 
and many of its properties remain to be studied in detail.

Since it is also the heaviest elementary particle known so far, its status in 
the SM is special. Indeed, the top quark is 40 times heavier than its weak
isospin partner the $b$ quark and its Yukawa coupling to the Higgs boson is close to 1.
The large top quark mass induces large contributions in virtual fermionic loops 
of radiative corrections. Since its decay time is smaller than the hadronization time, 
the top quark is the only quark that decays before hadronizing and so it can be used to probe 
the properties of a bare quark. All these characteristics may indicate that the
top quark could play a special role in the SM.

Top physics is a rich and still developping field in particular at the Tevatron, 
the proton-antiproton collider at Fermilab with a center of mass energy of 1.96~TeV,
which is currently the only place where the top quark can be produced directly.
Two multipurpose detectors are located around the Tevatron accelerator: CDF \dzero.
The Tevatron running period can be divided into three data taking periods: first the Run~I from
1993 to 1996 with a center of mass energy of 1.8~TeV and a delivered integrated luminosity
of $120~\pb$ per experiment, secondly the Run~IIa from 2002 to March 2006 
with an increased center of mass energy of 1.96~TeV and a delivered integrated luminosity of $1.5~\fb$ 
per experiment. Finally the third period from August 2006 is expected to last until 2009 or 2010 
with an expected delivered integrated luminosity from $6$ to $8~\fb$. The typical data taking 
efficiency of the experiments is above 85\%. The analyses presented in this article use
from $0.9$ to $2.2~\fb$ of data and are documented in \cite{d0,cdf}. The results are quoted as they
were presented at the time of the conference regardless on any later updates.

The top quark can be produced via two modes, either via the strong interaction leading to a production
by \ttbar\ pair or via the electroweak interaction leading to the single top production. The pair production
is the dominant mode at hadron colliders.

\section{Top Quark Pair Production: \ttbar\ Cross Sections}
\subsection{Top Quark Pair Production}
At Tevatron, the top quark pair production occurs by quark-antiquark annihilation in 85\% of the cases
while the remaining part is provided by gluon fusion.
The theoretical prediction for a top quark mass of $M_{top}=175$~GeV and the CTEQ6.5 PDF set~\cite{cteq} is:
$\sigma_{\ttbar} = 6.73~{\rm pb} \pm 10\%$ at Tevatron~\cite{cacciari} 
(it is $\sigma_{\ttbar} = 908~{\rm pb} \pm 10\%$ at LHC). 
Typically for $1~\fb$ of data, we expect to see around 300 \ttbar\ events in the lepton+jets channel per experiment at Tevatron.

Within the framework of the SM, the top quark decays approximately 100\% of the time into a $W$ 
boson and a $b$ quark. The \ttbar\ final states are classified according to the $W$ boson decay modes.
When the two $W$ bosons coming from the top and the antitop decay hadronically, the final state is called 
the alljets or full hadronic final state. This decay channel has the largest branching ratio but also suffers from the 
largest background coming from multijet events. The lepton+jets channel occurs 30\% of the time when one 
$W$ boson decays leptonically and one hadronically. This mode has reasonable statistics and reasonable
background. The two $W$ bosons decaying leptonically in 4\% of the cases lead to the dilepton channel that has 
low statistics but low background (here lepton denotes electron or muon). Finally in 20\% of the cases, the final state contains a $\tau$ 
which is challenging to identify in an hadronic environment.

\subsection{Motivations and Method}
The inclusive \ttbar\ cross section $\sigma_{\ttbar}$ is a quantity that allows to test the SM
by comparing the experimental measurements with the QCD Next-to-Leading Order (NLO) prediction. It also allows
to extract the top quark mass by comparing these two values. Measuring $\sigma_{\ttbar}$ enables as well as the possibility 
to probe for new physics that can manifest itself in anomalous \ttbar\ production rate or different cross section values
for different top decay channels. \ttbar\ events are also an important background for Higgs boson searches.

The \ttbar\ cross section is extracted using the following formula:
\begin{equation}
\sigma_{\ttbar} = \frac{N_{observed}-N_{background}}{\epsilon_{\ttbar}(M_{top}) L}
\end{equation}
where $N_{observed}$ and $N_{background}$ are the number of observed data events and predicted
number of background events. $N_{observed}-N_{background}$ can be evaluated either from a counting experiment 
or using a fit of a discriminating variable shape. $L$ is the recorded integrated luminosity that
is channel dependent and $\epsilon_{\ttbar}$ is the signal efficiency which is evaluated 
using \ttbar\ Monte Carlo (MC) samples correcting for any differences between data and MC.

\subsection{Cross Section Results in the Lepton+jets Channel}
The signature of \ttbar\ events in the lepton+jets channel consists of one high \pt\ isolated lepton,
large missing transverse energy (\met) and four jets among which two are $b$ quark jets. 
Tagging $b$ quark jets is very useful to select signal events.
Two main analysis strategies can be followed: either a topological approach
where a multivariable discriminant is formed using kinematic variables (aplanarity, sphericity, angles, invariant masses, ...) 
to separate signal from background or an approach based on b-tagging where b-tagging is used to enhance the sample purity.
The typical signal over background ratio is 2:3 for analysis without b-tagging and 4:1 in analysis 
with b-tagging requesting four jets. 
The main backgrounds in this channel come from $W$+jets events which is evaluated 
using MC but normalized to data before b-tagging and from multijet events where one jet fakes an isolated 
lepton. This later background is evaluated from data. Backgrounds coming from diboson, Z+jet and single top events 
are evaluated using MC.

Both \dzero\ and CDF measured this cross section with respectively $0.9~\fb$ and $1.12~\fb$ using b-tagging 
separating events depending on the lepton flavor, number of jets and number of b-tagged jets. The results are indicated in
Figure~\ref{fig:ttbar_xsection} and show a
12\% relative total error. The errors are already dominated by the systematic uncertainties. The dominant ones, except 
the luminosity error (6\%), come from the b-tagging uncertainty (6\%), MC modeling (4\%) and jet energy scale (4\%).

CDF also performed a measurement using soft lepton tagging. In the analysis, instead of using track impact parameters
or secondary vertex reconstruction to tag $b$ quarks, the semileptonic $b$ quark decay into either a soft electron or muon
is exploited. In this case the total cross section error is still dominated by the statistical uncertainty 
(Figure~\ref{fig:ttbar_xsection})
while the dominant source of systematic uncertainties comes from lepton tagging acceptance (8\% to 15\%) and fake rate (5\%).

Alternatively both \dzero\ and CDF performed a lepton+jets topological analysis where \dzero\ is using a likelihood
discriminant while CDF is using a neural network (NN). The discriminant output is fitted to extract the cross section 
(see Figure~\ref{fig:ttbar_xsection}). In these analyses, the statistical error is still important.
\dzero\ combines the topological with the b-tagging result that are 31\% correlated leading to a cross section 
measurement with 11\% total uncertainty.

The cross section results can be used to extract the top quark properties or to probe for new physics, particularly in the lepton+jets
channel which provides the most precise measurement. Comparing the combined topological and b-tagging \dzero\ cross section measurement described 
above with the computation from QCD, the top quark mass ($M_{top}$) can be extracted in a well defined renormalization scheme.
In order to do so, a joint likelihood is constructed resulting in the product of a theory likelihood 
(that takes both the factorization or renormalization scale uncertainties and the PDF error into account) times an experimental likelihood formed 
using the total experimental uncertainty as a function of $M_{top}$. The theory and experimental errors are assumed to be independent. 
The resulting top mass after integrating over the cross section is $M_{top} = 170 \pm 7$~GeV in agreement with the world average 
value~\cite{masscombi}.

The \ttbar\ cross section value can also be evaluated simultaneously with the ratio $R$ of $b$ quark to light quark branching ratios 
defined as:
\begin{equation}
R = \frac{Br(t \to Wb)}{Br(t \to Wq)} = \frac{|\vtb|^2}{|\vtb|^2 + |V_{ts}|^2 + |V_{td}|^2}
\end{equation}
where $q$ stands for any down type quark. Indeed the standard cross section measurements assume $R \approx 1$ as predicted by the SM.
A simultaneous extraction of $\sigma_{\ttbar}$ and $R$ allows to relax this assumption and to put a limit on the CKM matrix element
\vtb\ assuming three quark families and the matrix unitarity. \dzero\ performed this measurement and finds: $R = 0.97^{+0.09}_{-0.08}$~(stat+sys) 
and $|\vtb|>0.89$ at 95\% confidence level (CL).

\subsection{Cross Section Results in the Dilepton Channel}
The signature of \ttbar\ events in the dilepton channel consists of two high \pt\ isolated leptons,
large missing transverse energy (\met) and two $b$ quark jets.
The typical signal over background ratio is 3:1 without applying b-tagging. 
Complementary to the selection of two well identified leptons, one single lepton with an isolated track can also be requested
to enhance the signal acceptance. While b-tagging is not necessary to select a rather pure signal sample in the 
standard case, the lepton+track selection needs b-tagging to enhance the signal purity.
The main background in this channel comes from Drell-Yan production with fake \met. This background is 
estimated using data or MC. Multijet or $W$+jets events can also lead to the same final state when one or two
jets fake an isolated lepton. This type of background is estimated using data. Finally the diboson background
is evaluated using MC.

Both \dzero\ and CDF measured this cross section with respectively $1.1~\fb$ and $2~\fb$ without requesting  b-tagging.
The results are indicated in Figure~\ref{fig:ttbar_xsection} and show a
17\% relative total error. The total error is still dominated by the statistical uncertainty. The dominant systematic
uncertainties, except the luminosity error (6\%), come from the jet energy scale uncertainty (3\%) and MC normalization (3\%).

CDF also measured the dilepton cross section with b-tagging as well as the lepton+track cross section with and without
b-tagging (see Figure~\ref{fig:ttbar_xsection}).

Using both the dilepton cross section measurement described here and the lepton+jets one described in the previous section,
the $R_{\sigma}$ cross section ratio can be computed:
\begin{equation}
R_{\sigma} = \frac{\sigma(\ttbar \to l+jets)}{\sigma(\ttbar \to dilepton)}.
\end{equation}
The SM predicts $R_{\sigma}=1$ and any deviation from 1 is sensitive to $W$ disappearance: i.e. $t \to X b$.
For instance, in the minimal supersymmetric extension of the SM (MSSM), a charged Higgs boson can be lighter than the top quark and
the following decay can occur: $t \to Hb$ with $H^+ \to c \bar{s}$ which leads to $R_{\sigma}>1$.
\dzero\ combined its lepton+jets and dilepton cross section results to extract: 
$R_{\sigma} = 1.21^{+0.27}_{-0.26}$~(stat+sys) using a Feldman Cousins method~\cite{feldmancousins}.
Assuming a charged Higgs boson mass of $m_{H^\pm}=80$~GeV which decays exclusively into $c \bar{s}$, a limit on the branching ratio of 
top decaying to charged Higgs bosons can be set: $Br(t \to Hb)<0.35$ at 95\% CL.

Another interesting dilepton channel arises when one of the lepton is a $\tau$. Final states with $\tau$ leptons 
are challenging to identify in an hadronic environment but they are also the most sensitive channels to new physics.
Indeed for instance in the MSSM, when a charged Higgs boson is lighter than the top quark, it can also decay into a $\tau$: $H^+ \to \tau \nu$
therefore competing with the standard top decay into a $W$ boson.

At \dzero\ the identification of $\tau$ decaying hadronically is performed using three different NNs depending on the decay products 
of the $\tau$ classified in three cases: one charged pion or kaon and a neutrino, one  charged pion or kaon with a neutral pion and 
a neutrino or finally several charged pions or kaons and a neutrino.
\dzero\ performed a \ttbar\ cross section measurement in the lepton+$\tau$ channel using $1~\fb$ with b-tagging. The 
main background in 
this channel comes from $W$+jets events and is estimated from MC normalized to data. The multijet 
background where one jet fakes a $\tau$ and another one fakes an isolated lepton is estimated using same sign data events. 
The other backgrounds are evaluated using MC. The result is shown in Figure~\ref{fig:ttbar_xsection} and has a total relative 
uncertainty of 28 \% dominated by the statistical error.  The dominant systematic
uncertainties, except the luminosity error (6\%), come from background estimation (20\%), the b-tagging uncertainty (6\%) and from
the estimation of the $\tau$ fake rate (4\%).

\subsection{Cross Section Results in the Full Hadronic Channel}
The signature of \ttbar\ events in the full hadronic channel consists of six jets among which two are $b$ quark jets. 
The typical signal over background ratio is 1:2 before applying b-tagging. b-tagging is thus necessary to help
separating signal from background. With six jets in this channel, the combinatoric background is important.

CDF performed a \ttbar\ cross section measurement in this channel using $1~\fb$. The selection requires
from six to eight jets. The main background comes from multijet events estimated from data calibrating 
the tagging rate in the four jet bin which a background dominated sample. The signal from background separation is achieved using a NN that combines
jet invariant masses, sphericity, aplanarity, ... The result is shown in Figure~\ref{fig:ttbar_xsection}.
The total uncertainty of 25\% is dominated by systematic uncertainties where the jet energy scale error (16\%) and
the b-tagging uncertainty (7\%) are the dominant ones.

\subsection{Summary and Perspectives for Top Quark Pair Production}
\begin{figure}[htb]
\begin{center}
\epsfig{file=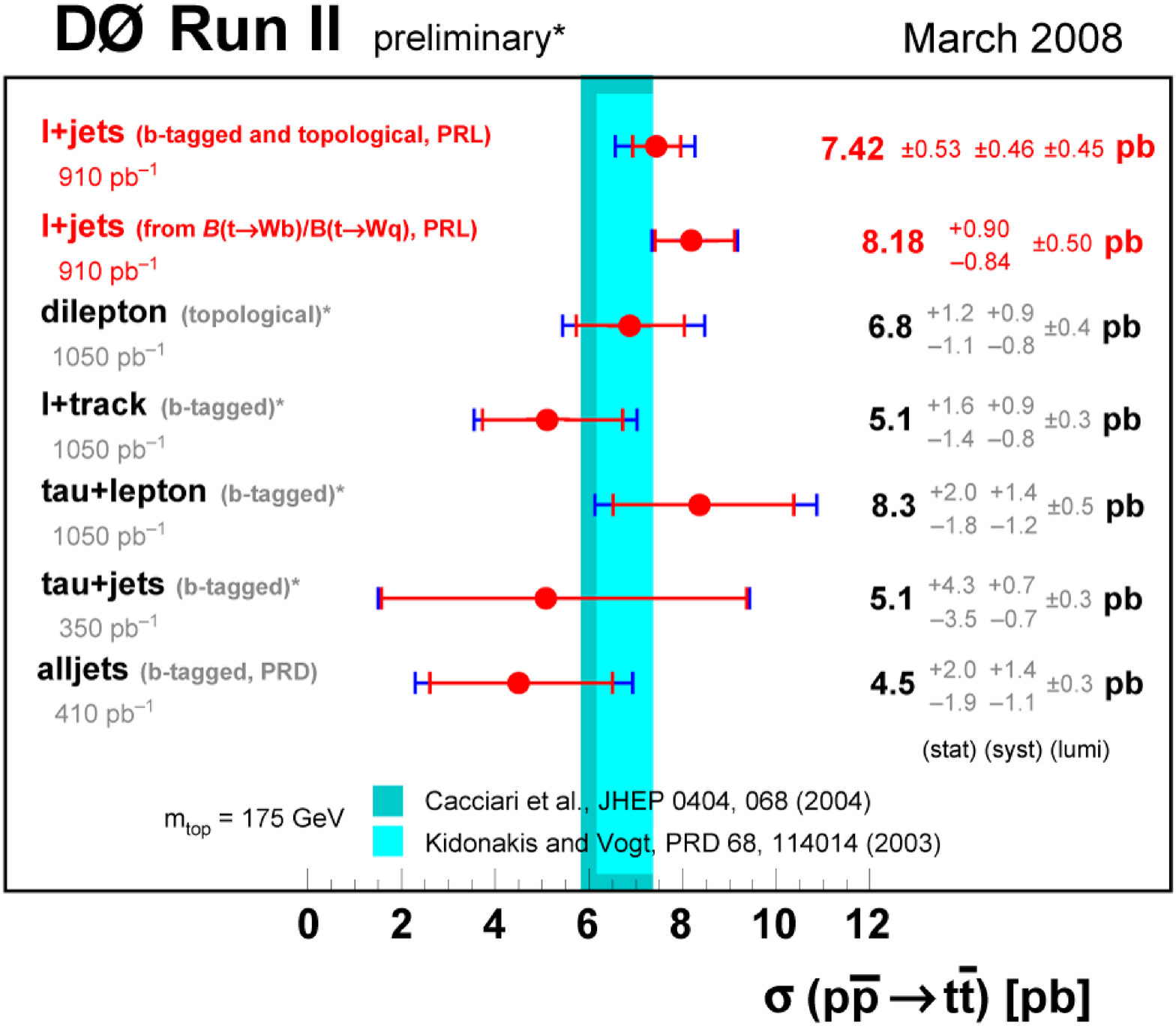,height=7cm}
\epsfig{file=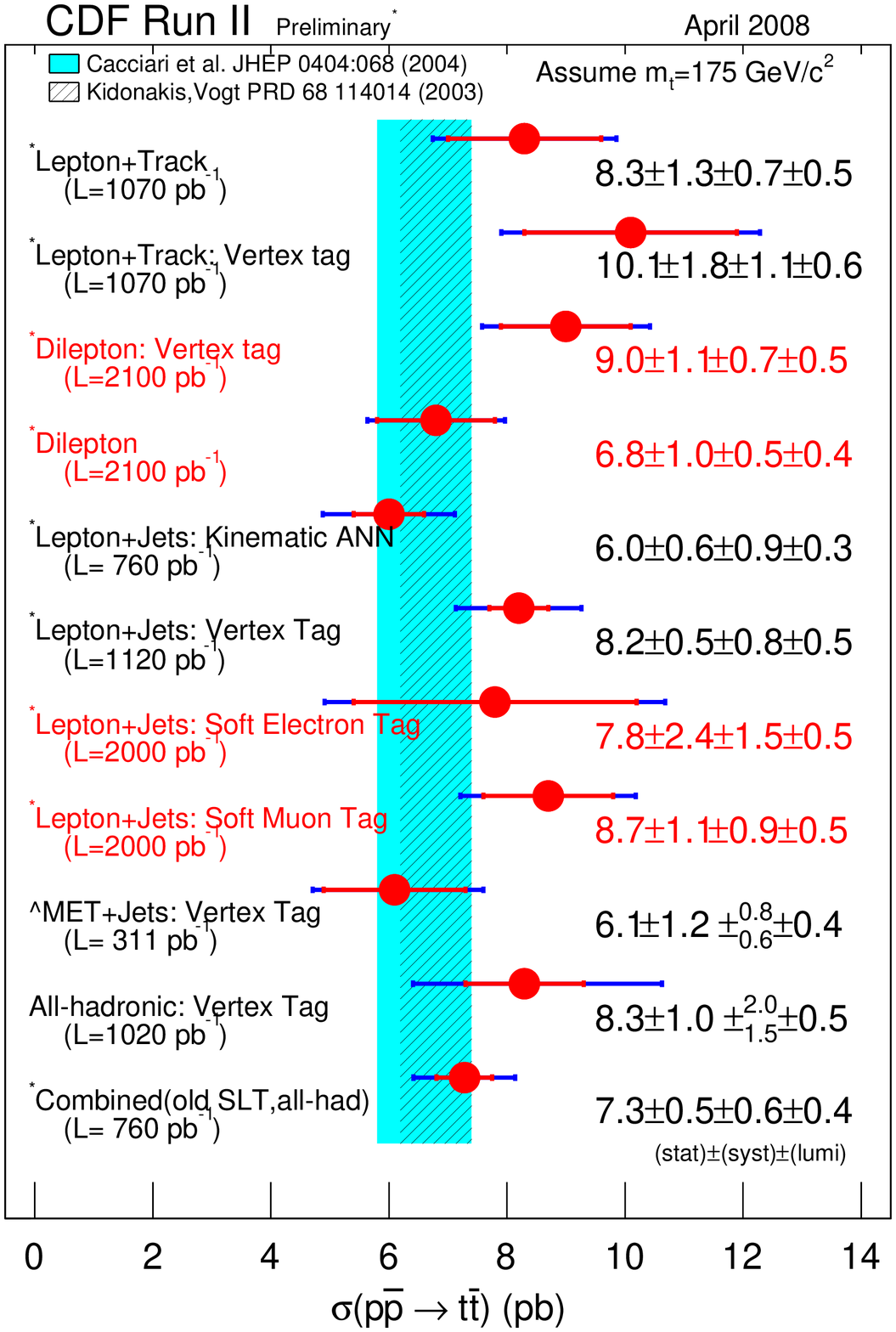,height=9cm}
\caption{Summary of the \dzero\ (left) and CDF (right) \ttbar\ cross section measurements.}
\label{fig:ttbar_xsection}
\end{center}
\end{figure}

All the \ttbar\ cross section measurements performed by the \dzero\ and CDF collaborations are summarized
in Figure~\ref{fig:ttbar_xsection}. All the measurements are in agreement with each other as well as with the theoritical
predictions. The current best single measurement shows a total uncertainty of 11\% which
is of the same order as the theoretical one. The systematic uncertainties dominate for the lepton+jets 
and full hadronic channels and future works will concentrate on reducing them. With the full luminosity that will be 
available at the end of Run~II, a 6\% total uncertainty could be achieved.
The expected uncertainty at the LHC using $10~\fb$ is from 5\% to 10\% dominated by the luminosity error.

\section{Single Top Quark Cross Sections}
\subsection{Production and Motivations}
At Tevatron, the top quark can be produced by electroweak processes via three types
of Feynman diagrams: the s-channel where a quark and an antiquark annihilate,
the t-channel where a light quark interacts with a $b$ quark from gluon splitting,
and the $W$ associated production where a gluon interacts with a light quark 
and produces a top quark in association with a $W$ boson.
At Tevatron, the main channel is the t-channel (65\%) followed by the s-channel (30\%)
while the W associated production is too small to be seen.
Since only one top quark is produced in the final state, only leptonic decay of the $W$ boson 
from the top quark can be considered. Thus the t-channel final state consists of 
one lepton, two $b$ quark jets including a very forward one and one light quark jet.
The s-channel final state consists of one lepton and two $b$ quark jets.
The combined s+t channel NLO theoretical prediction for a top quark mass of $M_{top}=175$~GeV and 
CTEQ5M1 PDF set~\cite{cteq} is:
$\sigma_t = 2.9~{\rm pb} \pm 14\%$ at Tevatron while it is $\sigma_t = 314~{\rm pb} \pm 20\%$ 
at LHC~\cite{sullivan}. Typically for $1~\fb$ of data, we expect to see around 50 single top events 
per experiment at Tevatron.

It is important to measure the single top cross section 
first because this mode is predicted by the SM, and the measured
cross section should be compared with the prediction to test the model. 
Since the single top cross section is directly proportional
to $|\vtb|^2$, it also allows a direct measurement of this CKM matrix element.
In addition this mode is sensitive to non standard processes like contribution from heavy $W'$ boson,
charged Higgs boson, flavour changing neutral currents (FCNC) or anomalous $Wtb$ coupling like $V+A$ contribution.
The single top production is also a background for Higgs boson searches since it leads to the same 
final state as the Higgs boson associated production $WH \to W b \bar{b}$.
However measuring the single top cross section is difficult since the cross section is low and 
suffers from high background from \ttbar, $W b \bar{b}$ and multijet events. Indeed 
after the preselection cuts, the typical signal over background ratio is 1:15. That's why
multivariate techniques need to be used to extract the signal.

\subsection{Results}
The preselection of single top events requires one isolated electron or muon with $\pt>15-20$~GeV,
large \met: $\met>15-15$~GeV and from two to four jets with $\pt>15-25$~GeV among which one or two
are b-tagged. 
The typical signal efficiency for the s-channel is around 3\% and for the t-channel around 2\%.

The diboson, Drell-Yan and \ttbar\ backgrounds are estimated using MC. The $W$+jets kinematics 
and flavor composition are evaluated using MC while its normalization comes from data before applying b-tagging.
The multijet background is evaluated from data either using sideband or non isolated lepton samples.

Both \dzero\ and CDF performed single top measurements using respectively $0.9~\fb$ and $2.2~\fb$.
The total number of selected data events by \dzero\ is 697 events in the 2 jet sample and 455 events in the 3 jet sample.
The total number of selected data events by CDF is 1535 in the 2 jet sample and 712 in the 3 jet sample.

Systematic uncertainties affect the normalization and/or the shapes of the distributions. The main normalization 
systematic errors come from \ttbar\ pairs and $W$+jets normalization uncertainty, jet energy scale and b-tagging 
uncertainties. The total cross section uncertainty is however dominated by the statistical error.

The \dzero\ collaboration already published a first $3 \sigma$ evidence for single top production~\cite{d0singletop}.
Three analyses are currently performed \cite{d0singletop2} that differ from the multivariate technique which is used after the 
preselection. \dzero\ uses methods based on boosted decision tree, on bayesian neural network and on matrix element.
The results are summarized in Table~\ref{tab:d0singletop}. \dzero\ uses the Best Linar Unbiased Estimate method \cite{blue} to combine 
the three analyses. The correlations amoung them are around 60\%. The combined results are also shown in table~\ref{tab:d0singletop}. 
\begin{table}[htb]
\begin{center}
\begin{tabular}{|l|c|c|c|c|}
\hline  
 &  Decision Tree &  NN & Matrix Element & Combined \\  \hline 
Expected significance & $2.1 \sigma$ & $2.2 \sigma$ & $1.9 \sigma$  & $2.3 \sigma$ \\
Observed significance & $3.4 \sigma$ & $3.1 \sigma$ & $3.2 \sigma$  & $3.6 \sigma$ \\
Cross section & $4.9^{+1.4}_{-1.4}$ pb & $4.4^{+1.6}_{-1.4}$ pb & $4.8^{+1.6}_{-1.4}$ pb & $4.7 \pm 1.3$ pb \\ \hline
\end{tabular}
\caption{\dzero\ single top results s and t channel combined using $0.9~\fb$ for $M_{top}=175$ GeV \cite{d0singletop2}.}
\label{tab:d0singletop}
\end{center}
\end{table}

The CDF collaboration uses a large dataset of $2.2~\fb$ and four analyses based on a likelihood function, neural network, matrix element and
boosted decision tree. The results are summarized in Table~\ref{tab:cdfsingletop}. CDF uses a neural network optimized using
neuro-evolution of augmentation topologies~\cite{neat} to combine the likelihood, neural network and matrix element analyses. 
The combined results are also shown in table~\ref{tab:cdfsingletop}.
\begin{table}[htb]
\begin{center}
\begin{tabular}{|l|c|c|c|c|c|}
\hline  
 &  Likelihood &  NN & ME & BDT & Combined \\  \hline 
Expected significance & $3.4 \sigma$ & $4.4 \sigma$ & $4.5 \sigma$ & $4.6 \sigma$ & $5.1 \sigma$ \\
Observed significance & $2.0 \sigma$ & $3.2 \sigma$ & $3.4 \sigma$ & $2.8 \sigma$ & $3.7 \sigma$ \\
Cross section & $1.8^{+0.9}_{-0.8}$ pb & $2.0^{+0.9}_{-0.8}$ pb & $2.2^{+0.8}_{-0.7}$ pb & $1.9^{+0.8}_{-0.7}$ pb & $2.2 \pm 0.7$ pb \\ \hline
\end{tabular}
\caption{CDF single top results s and t channel combined using $2.2~\fb$ for $M_{top}=175$ GeV.
ME stands for Matrix Element, and BDT for Boosted Decision Tree. 
The last column combines the likelihood, neural network and matrix element analyses.}
\label{tab:cdfsingletop}
\end{center}
\end{table}

\begin{figure}[htb]
\begin{center}
\epsfig{file=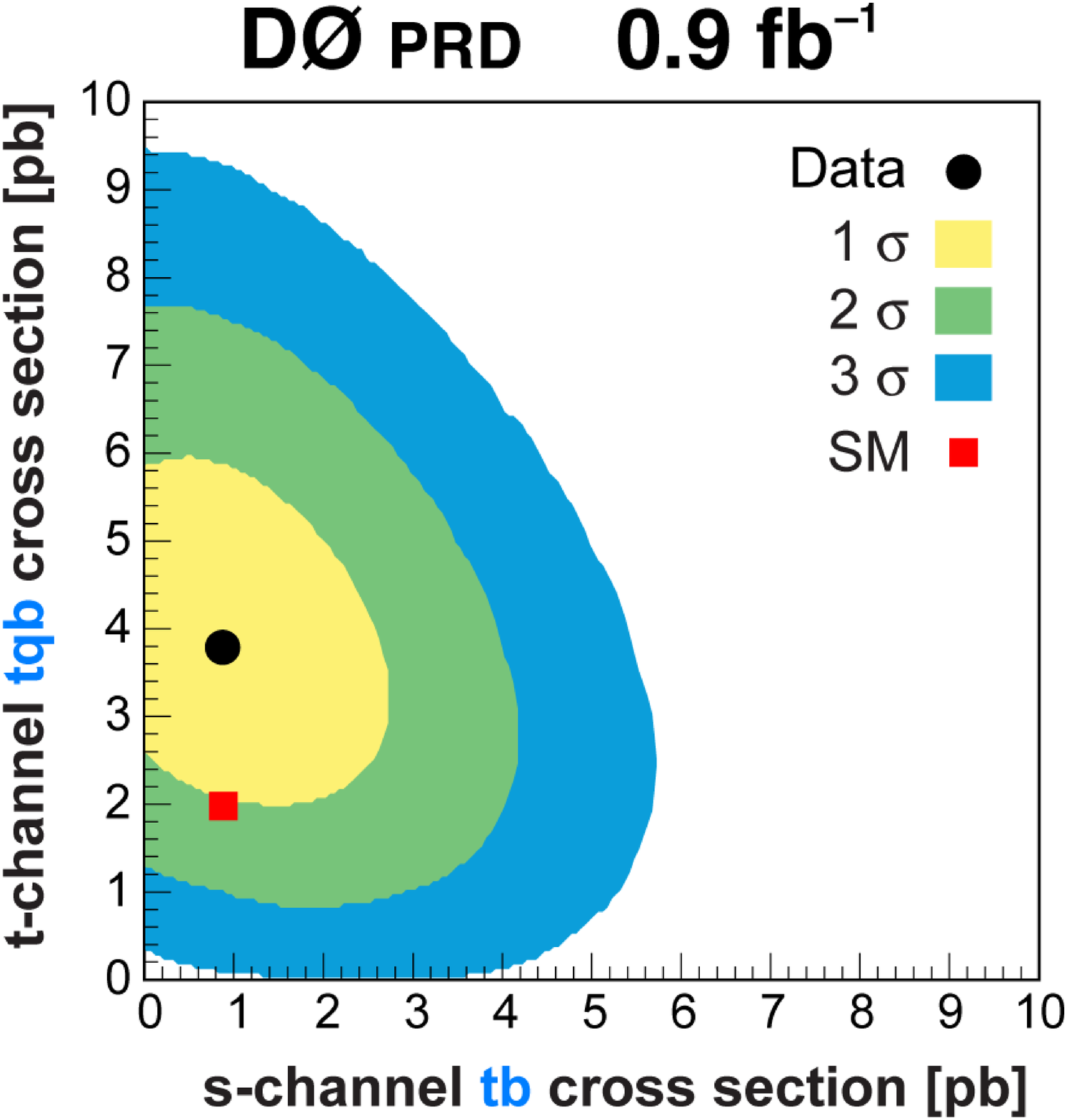,height=8cm}
\epsfig{file=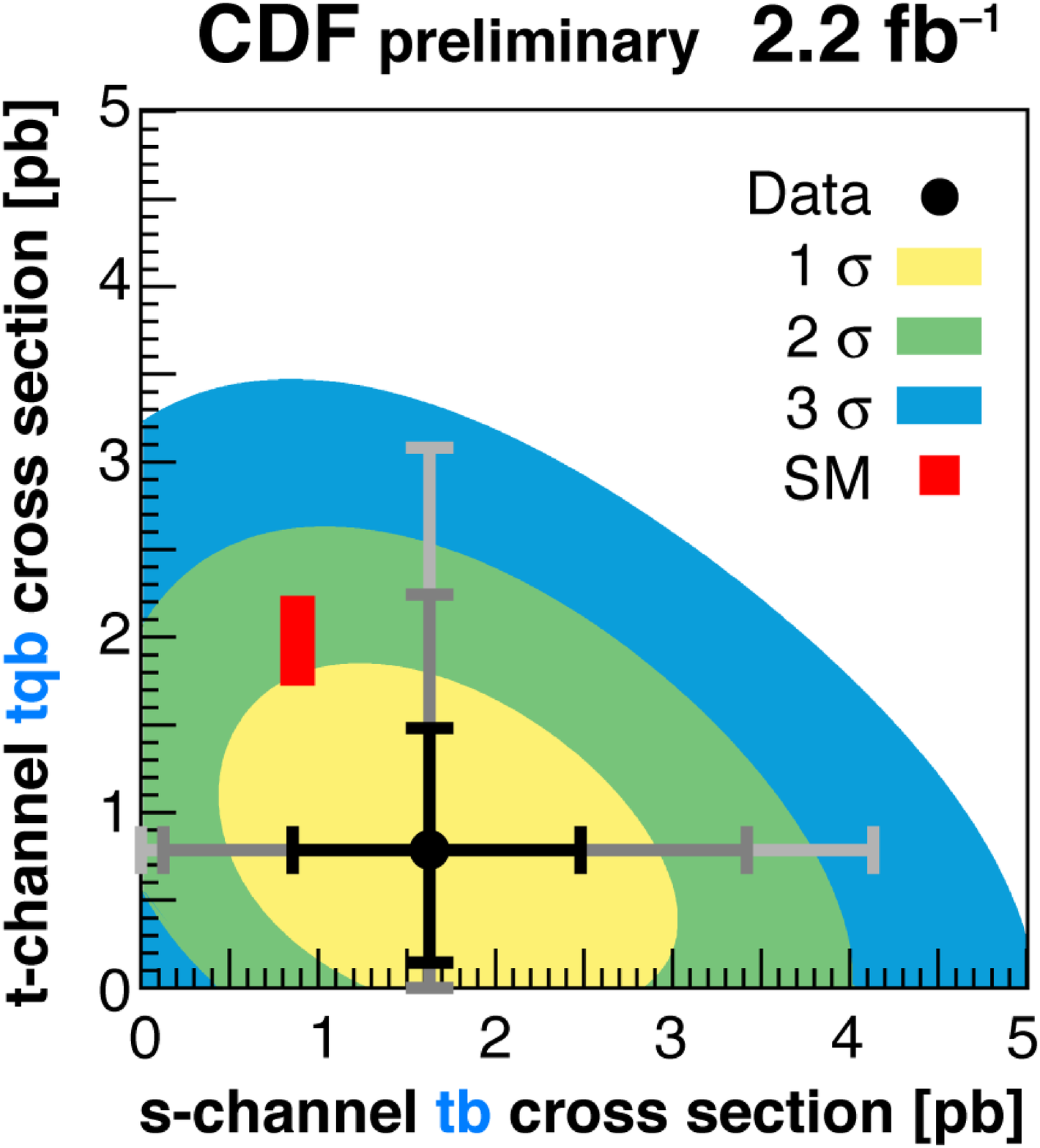,height=8cm}
\caption{\dzero\ (left) and CDF (right) s and t channel single top measurements}
\label{fig:singletop}
\end{center}
\end{figure}

Both \dzero\ and CDF also reported separate measurements of the s and t channels.
\dzero\ uses the s+t channel boosted decision tree where the s and t channel cross sections
are allowed to float. 
CDF uses neural network ouput templates for s and t channel separately and 
a 2-dimensional NN discriminant for the 2 jet 1 b-tag channel.
Hence \dzero\ measures: $\sigma(s-ch)=0.9$~pb and $\sigma(t-ch)=3.8$~pb.
CDF measures: $\sigma(s-ch)=1.6^{+0.9}_{-0.8}$~pb and $\sigma(t-ch)=0.8^{+0.7}_{-0.8}$~pb.
Figure~\ref{fig:singletop} shows these measurements with their uncertainties.

As already noticed, the single top cross section measurement allows to
measure directly the CKM matrix element $\vtb$ since it is proportional to $|\vtb|^2$.
To extract $\vtb$, several assumptions are made. First the single top production
is supposed to come exclusively from interaction with a $W$ boson (ie. no FCNC are 
considered), the top quark is assumed to decay as predicted by the SM 
(ie. $|V_{td}|^2 + |V_{ts}|^2 << |\vtb|^2$). Finally the $Wtb$ coupling is assumed to
be purely $V-A$ (ie. no anomalous $Wtb$ couplings). Under these hypotheses but without
any assumption regarding the number of quark families or CKM unitarity, the following lower 
bounds are extracted: $|\vtb| \ge 0.68$ at 95\% CL by \dzero\ and $|\vtb| \ge 0.66$ at 95\% CL 
by CDF both using a flat prior for \vtb\ from 0 to 1.

\subsection{Summary and Perspectives for Single Top Production}
\begin{figure}[htb]
\begin{center}
\epsfig{file=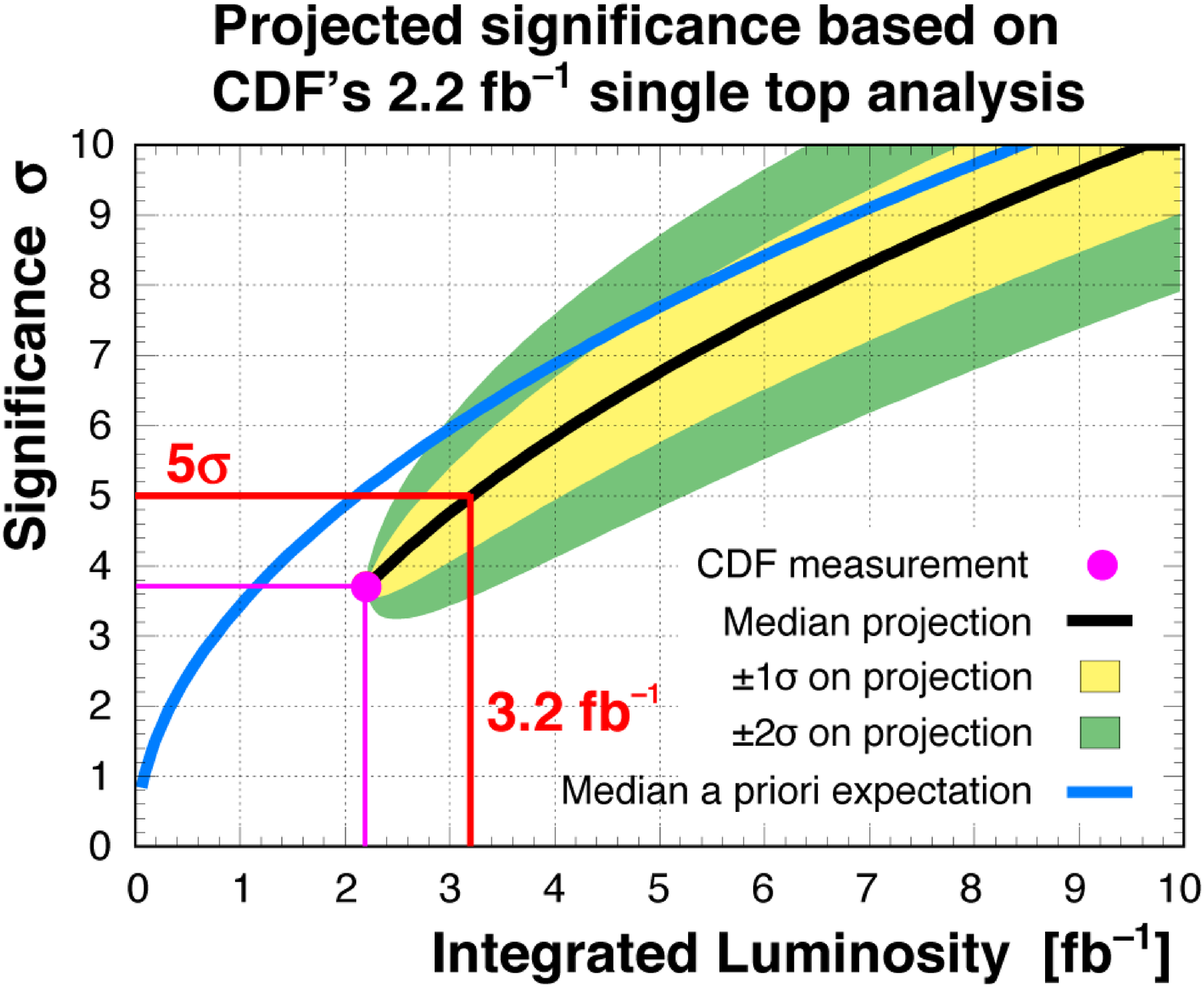,height=6cm}
\epsfig{file=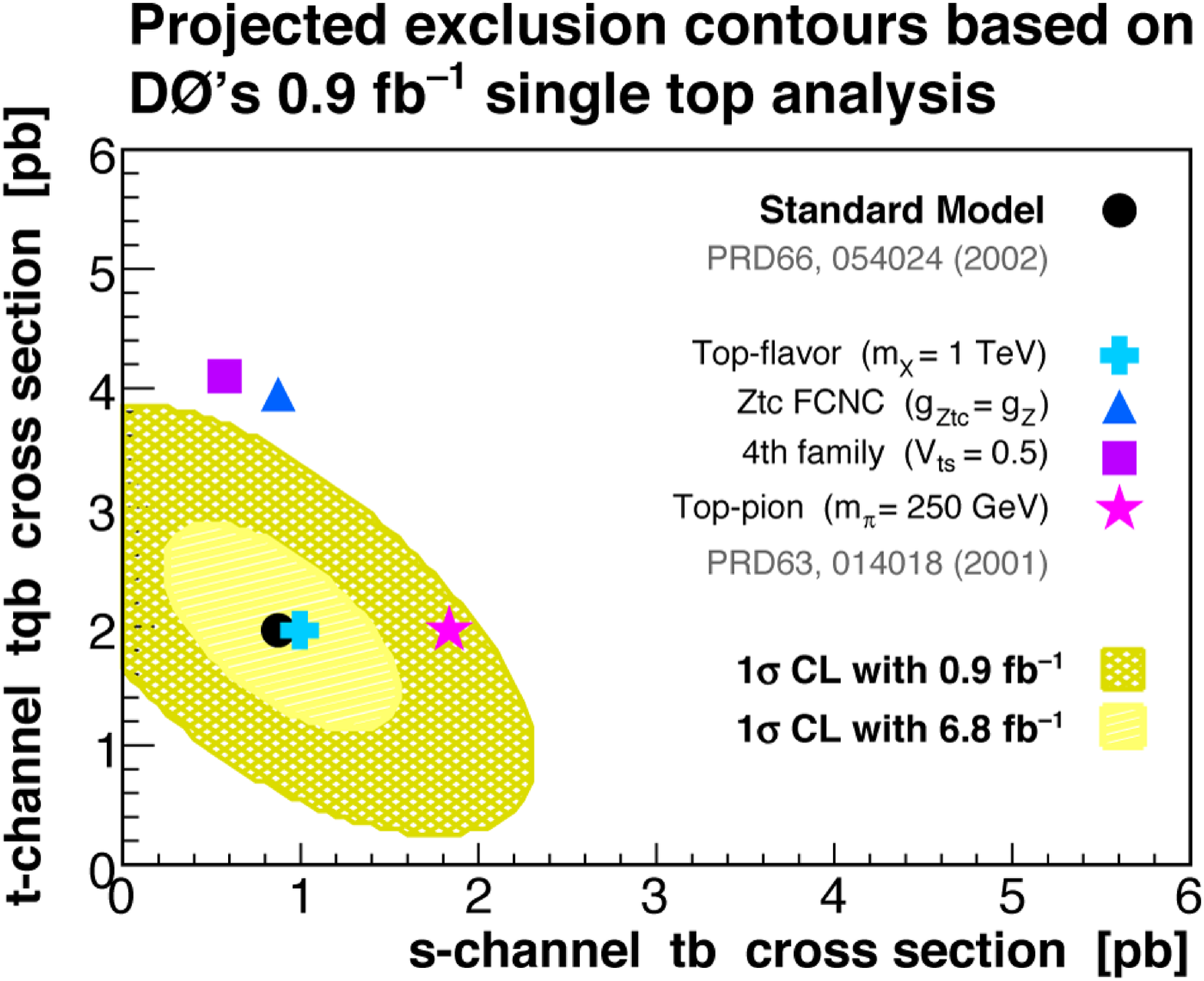,height=6cm}
\caption{CDF projected significance (left) and \dzero\ exclusion contours perspectives (right).}
\label{fig:perspectives}
\end{center}
\end{figure}

Even though single top cross section measurements are challenging, both \dzero\ and CDF
report $3 \sigma$ evidence for single top production (see Tables~\ref{tab:d0singletop} 
and \ref{tab:cdfsingletop}). By analyzing more statistics and based on the current analysis sensitivity, 
a $5 \sigma$ observation of this electroweak top quark production is expected soon (see Figure~\ref{fig:perspectives} left).
Analyzing the full run~II dataset will also allow to exclude an extended range of models beyond the SM that manifest themselves in the single top sector 
if no sign of new physics appears (see Figure~\ref{fig:perspectives} right).
LHC experiments will need around $10~\fb$ to see the t-channel and to discover the $W$ associated production channel. 
The s-channel will be more difficult to see at the LHC.

\section{Conclusion}
At the Run~II of the Tevatron, top physics entered a precision area. Indeed
the statistics is not the limiting factor anymore for the measurement of the  \ttbar\
cross section in the lepton+jets and full hadronic channels. The best measurement
currently achieves a 11\% total relative uncertainty which is of the same order as 
the theoretical prediction uncertainty. Because of the precision achieved, the SM
has been tested in the top quark sector by measuring the \ttbar\ cross section in 
all the possible channels and by reporting evidence for single top production.

The Tevatron will deliver from $6$ to $8~\fb$ of data by the end of 2009 or 2010.
At this time, even the dilepton \ttbar\ cross section will become limited by the 
systematic uncertainties. Future work to improve the \ttbar\ cross section
measurements will then focus on decreasing the systematic uncertainties.
At the end of Run~II, we also expect the discovery the single top in both 
the s and t channels. Of course with this statistics in hand, we can not
exclude to see surprises in the top quark physics sector.








 

\begin{thebibliography}{99}


\bibitem{topdiscovery}
  F.~Abe {\it et al.}  [CDF Collaboration],
  Phys.\ Rev.\ Lett.\  {\bf 74} (1995) 2626
  [arXiv:hep-ex/9503002];
  S.~Abachi {\it et al.}  [D0 Collaboration],
  Phys.\ Rev.\ Lett.\  {\bf 74} (1995) 2632
  [arXiv:hep-ex/9503003].

\bibitem{d0}
\verb|http://www-d0.fnal.gov/Run2Physics/top/top_public_web_pages/top_public.html|

\bibitem{cdf}
\verb|http://www-cdf.fnal.gov/physics/new/top/top.html|

\bibitem{cteq} 
 W.~K.~Tung, H.~L.~Lai, A.~Belyaev, J.~Pumplin, D.~Stump and C.~P.~Yuan,
  JHEP {\bf 0702} (2007) 053
  [arXiv:hep-ph/0611254];\\
  H.~L.~Lai, P.~Nadolsky, J.~Pumplin, D.~Stump, W.~K.~Tung and C.~P.~Yuan,
  JHEP {\bf 0704} (2007) 089
  [arXiv:hep-ph/0702268].

\bibitem{cacciari} 
  M.~Cacciari, S.~Frixione, M.~L.~Mangano, P.~Nason and G.~Ridolfi,
  JHEP {\bf 0809} (2008) 127
  [arXiv:0804.2800 [hep-ph]].

\bibitem{masscombi}
    [CDF Collaboration and D0 Collaboration],
  arXiv:0803.1683 [hep-ex].

\bibitem{feldmancousins}
  G.~J.~Feldman and R.~D.~Cousins,
  Phys.\ Rev.\  D {\bf 57} (1998) 3873
  [arXiv:physics/9711021].

\bibitem{sullivan}
  Z.~Sullivan,
  Phys.\ Rev.\  D {\bf 70} (2004) 114012
  [arXiv:hep-ph/0408049].

\bibitem{d0singletop}
  V.~M.~Abazov {\it et al.}  [D0 Collaboration],
  Phys.\ Rev.\ Lett.\  {\bf 98} (2007) 181802
  [arXiv:hep-ex/0612052].

\bibitem{d0singletop2}
  V.~M.~Abazov {\it et al.}  [D0 Collaboration],
  Phys.\ Rev.\  D {\bf 78} (2008) 012005
  [arXiv:0803.0739 [hep-ex]].

\bibitem{blue}
  L.~Lyons, D.~Gibaut and P.~Clifford,
  Nucl.\ Instrum.\ Meth.\  A {\bf 270} (1988) 110;\\
  A.~Valassi,
  Nucl.\ Instrum.\ Meth.\  A {\bf 500} (2003) 391.

\bibitem{neat}
  K. O. Stanley and Risto Miikkulainen, Evolutionary Computation, 
  {\bf 10 (2)} (2002) 99-127.

\end{thebibliography}
\end{document}